\documentclass[%
 reprint,superscriptaddress,
 amsmath,amssymb,
 aps,
]{revtex4-2}

\usepackage{array}
\usepackage{float}
\usepackage{graphicx} 
\usepackage{amsmath,amsfonts,amssymb,graphicx}
\usepackage[english]{babel}
\usepackage[utf8]{inputenc}
\usepackage[T1]{fontenc}
\usepackage{bbold}
\usepackage{caption,subcaption}
\usepackage{color}
\usepackage{algorithmic}
\usepackage{hyperref}
\usepackage[ruled,lined]{algorithm2e}
\usepackage{blindtext}
\usepackage[color=yellow]{todonotes}
\usepackage{booktabs}
\usepackage{multirow}
\usepackage{colortbl} 
\usepackage{xcolor}
\usepackage{tikz}
\usetikzlibrary{quantikz2}
\usepackage{dsfont}
\usepackage{wrapfig,tcolorbox,cleveref}

\newtcolorbox[auto counter,crefname={box}{boxes}]{pabox}[2][]{%
title=Box 1 Timeline,
colback=white,
colframe=black,
colbacktitle=white,
coltitle=black,
fonttitle=\bfseries,
title=Box ~\thetcbcounter. #2, label={#1}}

\makeatletter
\newcommand*{\balancecolsandclearpage}{%
  \close@column@grid
  \cleardoublepage
  \twocolumngrid
}
\makeatother

\begin{document}


\title[]{Estimation of trace distance between two arbitrary quantum states}

\author{Sanchita Ghosh}
\email{sanchita.ghosh14@gmail.com}
\affiliation{Faculty of Electrical Engineering, Mathematics and Computer Science, Delft University of Technology, Delft, 2628 CD, Netherlands.}
\affiliation{Center for Quantum Engineering, Research and Education (CQuERE), TCG CREST, Salt Lake, Sector 5, Kolkata 700091, India.}
\author{Anumita Mukhopadhyay}
\email{anumitamukherjee455@gmail.com}
\affiliation{Center for Quantum Engineering, Research and Education (CQuERE), TCG CREST, Salt Lake, Sector 5, Kolkata 700091, India.}
\affiliation{Academy of Scientific and Innovative Research (AcSIR), Ghaziabad- 201002, India.}
\author{Anindita Bera}
\email{aninditabera@bitmesra.ac.in}
\affiliation{Department of Mathematics, Birla Institute of Technology - Mesra, Jharkhand 835215, India.}
\author{Prasenjit Deb}
\email{devprasen@gmail.com}
\affiliation{Center for Quantum Engineering, Research and Education (CQuERE), TCG CREST, Salt Lake, Sector 5, Kolkata 700091, India.}
\author{Shibdas Roy}
\email{roy.shibdas@gmail.com}
\affiliation{Center for Quantum Engineering, Research and Education (CQuERE), TCG CREST, Salt Lake, Sector 5, Kolkata 700091, India.}
\affiliation{Academy of Scientific and Innovative Research (AcSIR), Ghaziabad- 201002, India.}

\begin{abstract}
When it comes to discriminating between two quantum states, trace distance is one of the well-known metrics used in quantum computation and quantum information theory. While there are several quantum algorithms for calculating the trace distance between two quantum states, computing it for any two general density matrices remains computationally demanding. In this paper, we propose a quantum algorithm based on the exponentiation of the density matrix and the improved quantum phase estimation (IQPE) to determine the trace distance for both pure and mixed states, with a time complexity of $O(N^2/\varepsilon^6)$ where $N$ is the number of qubits of the given states and $\varepsilon$ is the simulation or estimation precision error. We demonstrate its ability to predict the quantity with proof-of-principle simulations and also quantum hardware computations on the IBM quantum computers, confirming its promise for near-term quantum devices. 

\end{abstract}

\maketitle

\section{Introduction}
In quantum information theory, trace distance is a ubiquitous metric when it comes to distinguishing two quantum states. It is known to be the quantum version of Kolmogorov distance between two classical probability distributions \cite{nielsen}. It measures how close two quantum states are. The measure of trace distance has several applications and usage in quantum mechanics. Foundational tenets of quantum mechanics allow for discriminating between two mutually orthogonal states but non-orthogonal states cannot be discriminated perfectly, as no-cloning theorem \cite{WZ,DD,HPY} strictly forbids us from doing so. Additionally, the discrimination with certainty is not guaranteed even for multipartite orthogonal states, if only local operations and classical communication (LOCC) are allowed. Measurement of state distinguishability before and after an operation is key for evaluating algorithmic efficiency and fundamental constraints of a quantum hardware. Trace distance is a significant measure for comparing quantum processes \cite{gil}. It is also an indispensable tool for measuring non-Markovianity of quantum dynamics by measuring distinguishability of a pair of states through a quantum channel where increase in distinguishability indicates non-Markovian nature of a quantum evolution \cite{BLP}. The detection of the initial correlation between the system and the environment, trace distance is widely used by measuring the distinguishability between a pair of states \cite{amato,dajka,wissman,qcorrelation}. Apart from that, it is also used to quantify Bell non-locality \cite{bell}, to capture quantum coherence \cite{qcoherence}, in remote state preparation \cite{Ai-Xi} and in quantum networking to detect how a state changes through a quantum network \cite{Campbell,network}. 
The trace distance $D(\rho,\rho')$~\cite{nielsen} between the two given N-qubit states $\rho$ and $\rho'$ is defined as
\begin{equation}\label{eq:trdis}
    D(\rho,\rho') = \frac{1}{2}\mid\mid\rho - \rho'\mid\mid_1
\end{equation}
where $\mid \mid .\mid \mid_1$ is the trace norm as given below:
\begin{equation}
    \mid\mid\rho-\rho'\mid\mid_1 = {\rm Tr}\sqrt{(\rho-\rho')^\dagger(\rho-\rho')} = {\rm Tr}\sqrt{(\rho-\rho')^2}
\end{equation}

Note that if $\lambda_j$ are the eigenvalues of $\rho-\rho'$, which is Hermitian, then we can also define $D(\rho,\rho')$ as follows: 
\begin{equation}\label{eq:D eq lambda}
    D(\rho,\rho') = \frac{1}{2}\sum_{j=0}^{2^N-1}\mid\lambda_j\mid
\end{equation}
i.e. the trace distance between $\rho$ and $\rho'$ is half the sum of the absolute values of the eigenvalues of $\rho-\rho'$.
The value of trace distance is $0$ for two indistinguishable states and it is $1$ for two orthonormal states. Fidelity is another measure of closeness between two states and they are relatable via bounded inequalities. Trace distance \cite{CWH1,CWH2} and fidelity \cite{AU} are two of the most commonly employed distinguishability measures between quantum states. 
Compared to fidelity, the trace distance has an operational interpretation as the distinguishing advantage
in the optimal success probability when trying to distinguish two states that are chosen uniformly at random. In recent times,  several quantum algorithms \cite{LLSL,CSZW,WGLZY,WZ1,RASW} have been proposed for estimating these distinguishability measures. For instance, Ref.~\cite{LLSL} has presented three variational quantum algorithms on NISQ devices to estimate the trace norms for different situations. Ref.~\cite{CSZW} has introduced hybrid quantum–classical algorithms, namely, the variational trace distance estimation algorithm for these two distance measures on near-term quantum devices without any assumption of input state. Refs.~\cite{WGLZY,WZ1} show that estimating the trace distance of two quantum states is possible in quantum polynomial time when one of the states is low rank. In Ref.~\cite{qft}, trace distance is measured using path integral representation of reduced density matrix. 

In order to obtain the trace distance between two unknown quantum states, pure or mixed, it is necessary to characterize the states, which can be done using quantum state tomography \cite{nielsen,spintomo}. However, the requirement of resources increases exponentially as the system size increases. Thus, to avoid such demanding procedure, we exponentiate the density matrices using Lloyd-Mohseni-Rebentrost (LMR) algorithm as discussed in Box \ref{mybox1} and use them as unitaries in Improved Quantum Phase Estimation (IQPE) process as shown in the Box \ref{mybox} to estimate the eigenvalues. 
By estimating the eigenvalues of the difference of the two density matrices and summing their absolute values gives the trace distance.

\begin{pabox}[mybox1]{%
        Lloyd-Mohseni-Rebentrost (LMR) algorithm:}
        LMR algorithm is a powerful framework designed for the fault tolerant era, however its full potential is still not utilized. In this paper, we address the challenge of estimating trace distance by proposing a new quantum algorithm that leverages the fundamental components of the LMR framework.
        Density matrix exponentiation in LMR technique utilizes the SWAP gate, $S$, that acts as follows: 
\begin{eqnarray}
   S(\sigma \otimes \varsigma) S^\dagger &=& \varsigma \otimes \sigma,
\end{eqnarray}
The density matrix $\sigma$ is exponentiated in a register $A$ to obtain a unitary $e^{i\sigma t}$ and it acts on some state $\varsigma$ in another register $B$ having same dimension as $A$, as follows:
\begin{eqnarray}
   &&{\rm Tr}_A\left[e^{iSt}(\sigma \otimes \varsigma) e^{-iSt}\right] \\\nonumber
   &=&  {\rm Tr}_A\left[(e^{i\varsigma t}\sigma e^{-i\varsigma t})
   \otimes (e^{i\sigma t}\varsigma e^{-i\sigma t})\right] \\ 
   &=& e^{i\sigma t}\varsigma e^{-i\sigma t}.
\end{eqnarray}
This is attained by repetition of the following procedure \cite{LMR,KSGS}:
\begin{eqnarray}\label{eq:lmr}
{\rm Tr}_A\left[e^{iS\Delta t}(\sigma\otimes\varsigma) e^{-iS\Delta t}\right] = \varsigma - i\Delta t[\sigma,\varsigma]+O(\Delta t^2).
\end{eqnarray}
As the swap operator $S$ is sparse, it permits $e^{iS\Delta t}$ to be implemented efficiently \cite{BACS,HHL}. Furthermore, the total evolution time $t$ is given by $t=n\Delta t$, where $\Delta t$ is a small time interval and $n=O(t^2/\epsilon)$ is the required number of copies of $\sigma$, and so, the required number of times (\ref{eq:lmr}) must be repeated to simulate $e^{i\sigma t}$ with an error of $\epsilon$.
       
    \end{pabox}

    \begin{pabox}[mybox]{%
        Improved Quantum Phase Estimation (IQPE):}
        We begin with a two register system initialised as the state $|\chi_0\rangle|x_j\rangle$, where $|x_j\rangle$ is the $j$-th eigenstate of the Hermitian matrix $\mathbb{H}$ in the second register, that will be exponentiated, and in the first register $|\chi_0\rangle:=\sqrt{\frac{2}{T}}\sum_{\iota=0}^{T-1}\sin\frac{\pi(\iota+\frac{1}{2})}{T}|\iota\rangle$ for some large time $T$. The initial state $|\chi_0\rangle$ can be prepared with a margin of error $\epsilon_x$ in time ${\rm poly}\log_2(T/\epsilon_x)$ (see Section A of Supplementary material of Ref.~\cite{HHL}). 
        The conditional Hamiltonian evolution $\sum_{\iota=0}^{T-1}|\iota\rangle\langle\iota|\otimes e^{i\mathbb{H}\iota t_0/T}$ on the initial state in both registers is applied, followed by quantum Fourier transform (QFT) on the first register to obtain the state $\sum_{p=0}^{T-1}\mu_{p|j}|p\rangle|x_j\rangle$. We get the estimate $\tilde{y}_p$ of the $p$-th eigenvalue $y_p$ of $\mathbb{H}$ as $\tilde{y}_p:=\frac{2\pi p}{t_0}$, and relabel the Fourier basis states $|p\rangle$ to obtain $\sum_{p=0}^{T-1}\mu_{p|j}|\tilde{y}_p\rangle|x_j\rangle$. For perfect phase estimation, we have $\mu_{p|j}=1$ if $\tilde{y}_p=y_j$, and $0$ otherwise. Thus, we get the state $|\tilde{y}_j\rangle|x_j\rangle$, which in turn gives the estimate of $y_j$ upon measuring the first register. The error in the method is $\theta = O(1/t_0)$ in estimating $y_j$ \cite{HHL}, where $\theta/2$ is the error in trace distance (see just before Section A and just before Theorem 6 in the Supplementary material of Ref.~\cite{HHL}).
    \end{pabox}%
In Section \ref{sec:algo}, we discuss the steps of our algorithm elaborately. In Section \ref{sec:simulation}, we show the simulation of our algorithm and discuss about the hardware results. In Section \ref{sec:complex}, we calculate the complexity of our algorithm. In Section \ref{sec:discussion}, we discuss key insights about the algorithm, followed by Section \ref{sec:con}, where we conclude the paper.

\section{Algorithm for trace distance}\label{sec:algo}
 We now describe our algorithm to estimate the trace distance between two  $N$-qubit states $\rho$ and $\rho'$, given some identical copies of them. We first take $\vartheta:=(\mathbb{I}/2)\otimes\rho=\frac{1}{2}\left[\begin{array}{cc}\rho & 0\\0 & \rho\end{array}\right]$ and $\vartheta':=(\mathbb{I}/2)\otimes\rho'=\frac{1}{2}\left[\begin{array}{cc}\rho' & 0\\0 & \rho'\end{array}\right]$ to obtain an operator $\frac{\Omega}{2}:=\begin{pmatrix}\frac{1}{2}(\rho-\rho') & 0\\0 & \frac{1}{2}(-\rho+\rho')\end{pmatrix}$, which is a $2^{(N+1)} \times 2^{(N+1)}$ matrix. $\Omega/2$ is created by using LMR algorithm as in Box \ref{mybox1}, where an operator $e^{i\mathcal{T}t}$ instead of $e^{i\mathcal{S}t}$ is applied on $\vartheta$ and $\vartheta'$. Here $\mathcal{T}:=\mathcal{Z}\otimes\mathcal{S}=\left[\begin{array}{cc}\mathcal{S} & 0\\0 & -\mathcal{S}\end{array}\right]$, $\mathcal{S}$ is the Swap operator and $\mathcal{Z}$ is the $2\times 2$ Pauli-$Z$ unitary operator $\left[\begin{array}{cc}1 & 0\\0 & -1\end{array}\right]$. Thus, the dimension of $\mathcal{T}$ matches with the dimension of $\Omega/2$. To obtain exponentiated form of $\Omega/2$ we have taken $\vartheta$, $\vartheta'$ and an ancilla $\Sigma$, which is a density matrix of the same dimension as $\Omega/2$, i.e.\ $2^{N+1}\times 2^{N+1}$, in three registers as $\chi= \vartheta \otimes \vartheta' \otimes \Sigma$. The operator $e^{i \mathcal{T}\Delta t}$ acts on registers 1 and 3 and $e^{-i \mathcal{T}\Delta t}$ acts on registers 2 and 3 as follows:
\begin{eqnarray}
    \chi_2&=& e^{-i \mathcal{T}_{2,3}\Delta t}e^{i \mathcal{T}_{1,3}\Delta t}\chi e^{-i \mathcal{T}_{1,3}\Delta t}e^{i \mathcal{T}_{2,3}\Delta t} \\ \nonumber
    &=& e^{-i \mathcal{T}_{2,3} \Delta t}(\chi + i\Delta t [\mathcal{T}_{1,3},\chi])e^{i \mathcal{T}_{2,3}\Delta t}\\ \nonumber
    &\approx& (\chi + i\Delta t [\mathcal{T}_{1,3},\chi])- i\Delta t [\mathcal{T}_{2,3},(\chi + i\Delta t [\mathcal{T}_{1,3},\chi])]\\
    &\approx& \chi + i \Delta t [\mathcal{T}_{1,3},\chi] - i\Delta t [\mathcal{T}_{2,3}, \chi]. 
\end{eqnarray}
In order to get $\chi_2$, we have implemented Baker-Campbell-Hausdorff (BCH) expansion and taken first-order approximation of the expansion. 
Now tracing out first two registers from $\chi_2$ we get:
\begin{eqnarray}
    {\rm Tr}_{1,2}[\chi_2]&=& \Sigma + i \Delta t {\rm Tr}_1[\mathcal{T}_{1,3}, \vartheta \otimes \Sigma]\\ \nonumber
    && - i\Delta t {\rm Tr}_{2}[\mathcal{T}_{2,3}, \vartheta' \otimes \Sigma]\\ \nonumber
    &\approx& \Sigma + i \Delta t \left[\frac{\Omega}{2},\Sigma \right]+ O(\Delta t^2)\\
    &\approx& e^{i \frac{\Omega}{2} \Delta t} \Sigma e^{-i \frac{\Omega}{2} \Delta t} + O(\Delta t^2)
\end{eqnarray}
Thus, $\frac{\Omega}{2}$ is exponentiated by repeating the above step for $n$ times, following the same procedure as given in box (\ref{mybox1}). From exponentiated $\Omega/2$, we have created controlled-unitary, $U_{AB}$ following Ref.~\cite{LMR}. The form of $\Omega/2$ is to ensure that its eigenvalues will  always come in pairs of zeros or pairs of equal, positive and negative values. For every eigenvalue $\lambda_j:=-\varpi_j$ and $\varpi_j=\mid\lambda_j\mid$ of matrix $\Omega/2$, there would be another eigenvalue $\lambda_m:=\varpi_j$ with $m\neq j$.

We now implement the above-mentioned technique using which we estimate the trace distance, extracting three quantities, which we call $\kappa_1$, $\kappa_2$, and $\ell$ where $\kappa_1$ and $\kappa_2$ are two auxiliary scalar quantities, each obtained from a dedicated phase-estimation circuit run with a different choice of $U_{AB}$, and $\ell$ is an auxiliary quantity related to the number of zero eigenvalues of $\Omega/2$; Eq.~(\ref{eq:kappa2-kappa1+l_estm}) below shows how the three are combined algebraically to yield the trace-distance estimate $\tilde{D}(\rho,\rho')$.

\begin{enumerate}
    \item \textit{Initial state:} We start with two quantum registers - the clock register, initialized in an $M$-qubit state $|\Psi_0\rangle$, the state described in improved quantum phase estimation \cite{HHL} and the input state register initialized to a maximally mixed state, $\rho_{\text{in}}=\mathbb{I}/2^{N+1}$, conveniently written as $\mathbb{I}/2^{N+1} = (1/2^{N+1})\sum_{k=0}^{2^{N+1}-1}|\xi_k\rangle\langle\xi_k|$, where $\{|\xi_k\rangle\} $ are the eigenstates of the unitary $U_{AB}$. The unitary $U_{AB}$ will be defined soon for $\kappa_1$ and $\kappa_2$.
    
    Overall, the initial state with the two registers can be described as: \begin{equation}\label{eq:initial state}
    \frac{1}{2^{N+1}}\sum_{k=0}^{2^{N+1}-1}|\Psi_0\rangle\langle \Psi_0|\otimes|\xi_k\rangle\langle\xi_k|.
    \end{equation}

    \item \textit{First Quantum Phase Estimation (QPE):} Now we perform an Improved QPE with the input state $\rho_{\text{in}}=\mathbb{I}/2^{N+1}$ and unitary $U_{AB}$, as illustrated in Fig.~\ref{fig:1},to obtain the following output \begin{equation}\label{eq:trace-alg-qpe1}
    \frac{1}{2^{N+1}}\sum_{k=0}^{2^{N+1}-1}|\tilde{\lambda}_k\rangle\langle\tilde{\lambda}_k|\otimes|\xi_k\rangle\langle\xi_k|.
    \end{equation}
   Here $|\tilde{\lambda}_k\rangle$ are the estimated eigenvalues corresponding to eigenstates $|\xi_k\rangle$.
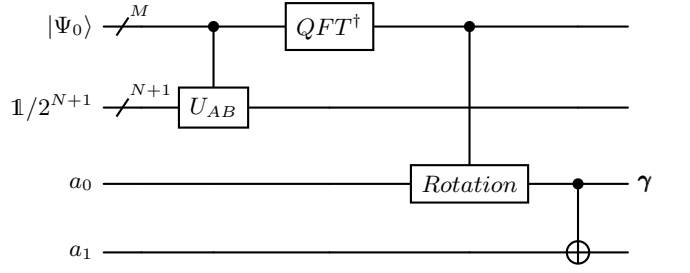
\begin{figure}
    \centering
        \begin{quantikz}
        \lstick{$|\Psi_0\rangle$} & \qwbundle{M} & \ctrl[vertical wire= q]{1}   & \gate{QFT^{\dagger}} & \ctrl{2} &&\\
        \lstick{$\mathbb{I}/2^{N+1}$} & \qwbundle{N+1} &                     \gate{U_{AB}} &&&& \\
        \lstick{$a_0$} &  &                    &&          \gate{Rotation} & \ctrl{1}  &  \rstick{$\pmb{\gamma}$} \\
        \lstick{$a_1$} &  &                    &&           &              \targ{} &\\
        \end{quantikz}
    \caption{First Quantum Phase Estimation (QPE) and Controlled Rotation}
    \label{fig:1}
\end{figure}

    \item \textit{Controlled Rotation:} Next we add an ancilla qubit, and perform a controlled rotation as it is done in the HHL algorithm in Ref.~\cite{HHL}, denoted by $a_0$ in Fig.~\ref{fig:1}. $a_0$ is initialized to $|0\rangle$ and rotated conditioned on $|\tilde{\lambda}_k\rangle$ to obtain $|\gamma_k\rangle =\sqrt{(1-\tilde{\lambda}_k^2)}|0\rangle+\tilde{\lambda}_k|1\rangle$. The overall state after this step is therefore
    \begin{equation}
    \frac{1}{2^{N+1}}\sum_{k=0}^{2^{N+1}-1}|\tilde{\lambda}_k\rangle\langle\tilde{\lambda}_k|\otimes|\xi_k\rangle\langle\xi_k|\otimes |\gamma_k\rangle\langle\gamma_k|.
    \end{equation}
Note that the decimal values that $\tilde{\lambda}_k$s take are $\{0.75,0.5,0.25,.125,\ldots\}$, depending on the number of clock qubits. 

    \item \textit{Create mixed state from rotated ancilla:} We again add an ancilla qubit $a_1$, initialized to $|0\rangle$, to the circuit as shown in Fig.~\ref{fig:1}. This is followed by the application of a CNOT gate with the target on $a_1$ and control on the $a_0$ qubit.
    
    After tracing out the recently added ancilla qubit $a_1$, we get the following state in the remaining registers 
    \begin{eqnarray}
    \frac{1}{2^{N+1}}\sum_{k=0}^{2^{N+1}-1}|\tilde{\lambda}_k\rangle\langle\tilde{\lambda}_k|\otimes|\xi_k\rangle\langle\xi_k|\otimes \nonumber  \\  \left[(1-\tilde{\lambda}_k^2)|0\rangle\langle 0|+\tilde{\lambda}_k^2|1\rangle\langle 1|\right].
    \end{eqnarray}


    \item \textit{Second Quantum Phase Estimation (QPE): } As in Ref.~\cite{HHL}, we uncompute the first register $|\tilde{\lambda}_k\rangle$ by undoing the phase estimation. Then we trace out the second register $|\xi_k\rangle$, to be effectively left with the following state, which we shall denote as $\gamma$. We simply perform another QPE considering the last register as follows. The density matrix from the last register is obtained as:

    \begin{equation}
    \gamma = \frac{1}{2^{N+1}}\sum_{k=0}^{2^{N+1}-1}\left[(1-\tilde{\lambda}_k^2) |0\rangle\langle 0|+\tilde{\lambda}_k^2|1\rangle\langle 1|\right]
    \end{equation}
    With $|1\rangle$ as the input and the unitary as $W:=e^{i\gamma t_0}$,  obtained using the LMR technique~\cite{LMR} we perform a QPE, with a P-qubit $|\Psi_0\rangle$, to obtain the eigenvalue,
     $\Lambda_1 :=\frac{1}{2^{N+1}}\sum_k\tilde{\lambda}_k^2$, as shown in Fig.~(\ref{fig:2}) and Fig.~(\ref{fig:3}).

\end{enumerate}

\subsubsection{Calculating $\kappa_1$, $\kappa_2$ and $\ell$}

The state in the clock register after the second QPE is $\tilde{\Lambda}_1 \approx \Lambda_1/4 =\frac{1}{2^{N+1}}\sum_k\tilde{\lambda}_k^2/4$, if we take the time  variables, $t$, $t_0$, for the two QPEs in the algorithm as $t = t_0 = \pi/2$. This is the result that is of interest to us and it changes based on the unitary $U_{AB}$ used in the First QPE. 

\paragraph{Calculating $\kappa_1$:}
Let $U_{AB} = U_A\cdot U_B = e^{i (\Omega/2) t} + O(t^2)$, where $O(t^2)$ is the Trotter error arising from approximating $e^{i (\vartheta-\vartheta') t}$ as $e^{i\vartheta t}\cdot e^{-i\vartheta' t}=U_A\cdot U_B$. Here we take $\vartheta=\mathbb{I}/2\otimes\rho$ and $\vartheta' =\mathbb{I}/2\otimes\rho'$. Then we define, $\frac{1}{2^{N+1}}\sum_k\frac{\tilde{\lambda}_k^2}{4} := \kappa_1$, which, in turn, yields: 
\begin{eqnarray}\label{eq:kappa1}
    \kappa_1 &=& \frac{1}{2^{N+1}}\sum_{k=0}^{2^{N+1}-1}\frac{\tilde{\lambda}_k^2}{4}\nonumber \\ \nonumber
    &=& \frac{1}{4\cdot 2^{N+1}}\sum\limits_{\substack{l=0\\\tilde{\varpi}_l\neq 0}}^{2^N-1}\left[\left(1-\frac{\tilde{\varpi}_l}{4}\right)^2+\left(\frac{\tilde{\varpi}_l}{4}\right)^2\right] \nonumber \\ \nonumber
    &&+ \frac{1}{4\cdot 2^{N+1}}\sum\limits_{\substack{l=0\\\tilde{\varpi}_l=0}}^{2^N-1}\left[\left(\frac{\tilde{\varpi}_l}{4}\right)^2+\left(\frac{\tilde{\varpi}_l}{4}\right)^2\right]\nonumber \\ \nonumber
    &=& \frac{1}{2^{N+3}}\sum\limits_{\substack{l=0\\\tilde{\varpi}_l\neq 0}}^{2^N-1}\left(1-\frac{\tilde{\varpi}_l}{2} +\frac{\tilde{\varpi}_l^2}{8}\right)\nonumber \\ \nonumber
    &&+ \frac{1}{2^{N+3}}\sum\limits_{\substack{l=0\\\tilde{\varpi}_l=0}}^{2^N-1}\frac{\tilde{\varpi}_l^2}{8}\nonumber \\ 
    &=& \frac{1}{2^{N+3}}\sum\limits_{\substack{l=0\\\tilde{\varpi}_l\neq 0}}^{2^N-1}\left(1-\frac{\tilde{\varpi}_l}{2} +\frac{\tilde{\varpi}_l^2}{8}\right).
\end{eqnarray}

\paragraph{Calculating $\kappa_2$:} 
Similarly, if we take $U_{AB} =e^{i(\mathbb{I}+\Omega/2)t}+O(t^2)$, where $O(t^2)$ is the Trotter error, then $\frac{1}{2^{N+1}}\sum_k\frac{\tilde{\lambda}_k^2}{4} = \kappa_2$, which, in turn, yields:

\begin{eqnarray}\label{eq:kappa2}
    \kappa_2 &=& \frac{1}{2^{N+1}}\sum_{k=0}^{2^{N+1}-1}\frac{\tilde{\lambda}_k^2}{4} \nonumber\\ \nonumber 
    &=& \frac{1}{4\cdot 2^{N+1}}\sum_{l=0}^{2^N-1}\left[\left(\frac{1-\tilde{\varpi}_l}{4}\right)^2+\left(\frac{1+\tilde{\varpi}_l}{4}\right)^2\right]\nonumber\\ 
    &=& \frac{1}{2^{N+3}}\left(\sum_{l=0}^{2^N-1}\frac{1}{8}+\sum_{l=0}^{2^N-1}\frac{\tilde{\varpi}_l^2}{8}\right).
\end{eqnarray}
\begin{figure}[t]
    \centering
    \begin{quantikz}
        \lstick{$|\Psi_0\rangle$} & \qwbundle{P} &\ctrl[vertical wire= q]{1}   & \gate{QFT^{\dagger}}  && \rstick{$|\tilde{\Lambda}_1\rangle$}\\
        \lstick{$|1\rangle$} &  &  \gate[2]{W} &&& \\
        \lstick{$\pmb{\gamma}^{\otimes Q}$} & \qwbundle{Q}  &  & && \\
    \end{quantikz}
    \caption{Second QPE}
    \label{fig:3}
\end{figure}
\begin{figure}
    \centering
        \begin{quantikz}
            \lstick{$|1\rangle$} && \swap{1}  \gategroup[wires=5, steps=4, style={dotted, rounded corners, inner sep=7pt}]{$W =e^{i\gamma t_0}$} & \swap{2} & \ldots & \swap{4} & & \rstick{$W|1\rangle\langle 1|W^{\dagger}$}\\
            \lstick{$\pmb{\gamma}$} &&    \targX{} &          & \ldots &          &\\
            \lstick{$\pmb{\gamma}$} &&     &       \targX{}   &  \ldots &          &\\
            \lstick{$\vdots$}  \\
            \lstick{$\pmb{\gamma}$} &&     &       &        \ldots & \targX{}   &
        \end{quantikz}
    \caption{Exponentiate $\gamma$. All the gates here are the exponentiated SWAP gate $e^{i\mathcal{S}\delta}$, where $\mathcal{S}$ denotes the SWAP gate and $\delta$ is the angle.}
    \label{fig:2}
\end{figure}

\paragraph{Calculating $\ell$:}
Combining (\ref{eq:kappa1}) and (\ref{eq:kappa2}) allows us to extract the trace distance. However, notice that
 if the matrix $\Omega/2$ is not full-rank, then there would be at least one pair of zero eigenvalues i.e. $\varpi_l=0$. Both (\ref{eq:kappa1}) and (\ref{eq:kappa2}) need this information about the number of eigenvalues that are zero, i.e, $\varpi_l=0$ and non-zero, i.e, $\varpi_l \neq 0$ in the matrix $\Omega/2$. To obtain the number of eigenvalues that are zero we now perform a QPE evolving the output of the first register given in (\ref{eq:trace-alg-qpe1})  using the LMR~\cite{LMR} technique to obtain the unitary $e^{i \beta t}$, where $\beta$ is given as:
 \begin{equation}\label{eq:beta}
\beta:=(1/2^{N+1})\sum_{k=0}^{2^{N+1}-1}|\tilde{\lambda}_k\rangle\langle\tilde{\lambda}_k|
\end{equation}
 We feed $|0\rangle$ as the input to this QPE, leading to an output $L$, as given below,
 \begin{equation}\label{eq:zeroeigen}
 L=\tilde{\ell}/(4\cdot 2^{N+1})=\tilde{\ell}/2^{N+3}.
 \end{equation}
 Here we denote by $\ell$ the true (integer) number of zero eigenvalues of $\Omega/2$, and by $\tilde{\ell}$ its phase-estimation-based \emph{estimate}, which is generally non-integer owing to the finite clock-qubit resolution of Box~\ref{mybox}; this is the quantity actually delivered by the circuit above and used throughout the remainder of this Section, and it is what is tabulated as $\tilde{\ell}$ in Table~\ref{table:hw results}.
From (\ref{eq:zeroeigen}), the number of eigenvalues that are zero, can be obtained as $\tilde{\ell}$, given as,
\begin{equation}
    \tilde{\ell} = L (2^{N+3}).
\end{equation}
The number of non-zero eigenvalues are then 
\begin{equation}\label{eq:l1}
    2^{N+1}-\tilde{\ell}
\end{equation}

\paragraph{Putting it all together:}

Combining (\ref{eq:kappa1}), (\ref{eq:kappa2}) and (\ref{eq:l1}), we obtain the following,

\begin{equation}
 2^{N+4}(\kappa_2 -\kappa_1)= 2\left[\sum_{l=0}^{2^N-1} 1/8 - \sum\limits_{\substack{l=0\\\tilde{\varpi}_l\neq 0}}^{2^N-1}1\right] +\sum_{l=0}^{2^N-1}\tilde{\varpi_l}
 \end{equation}
Note that there are about $1/2$ the total number of non-zero eigenvalues (that are negative) contributing to $\sum_{l=0, \tilde{\varpi}_l\neq 0}^{2^N-1}1$.
 \begin{equation}\label{eq:k2-k1}
2^{N+4}(\kappa_2 -\kappa_1)=2\left[\frac{2^N}{8}- \frac{2^{N+1}-\tilde{\ell}}{2}\right] +\sum_{l=0}^{2^N-1}\tilde{\varpi_l}  
\end{equation}
From (\ref{eq:k2-k1}), it is evident that
\begin{equation}
   \sum_{l=0}^{2^N-1}\tilde{\varpi_l}
= 2^{N+4}(\kappa_2 -\kappa_1)-\tilde{\ell}+\frac{7}{4}2^N.  
\end{equation}

From the above equation we get;

\begin{equation}\label{eq:kappa2-kappa1+l_estm}
     \begin{aligned}
        \frac{1}{2}\sum_{l=0}^{2^N-1}2\tilde{\varpi}_l \approx
         2^{N+4}(\kappa_2 - \kappa_1) + 7\cdot 2^{N-2} - \tilde{\ell} = \tilde{D}(\rho,\rho')
     \end{aligned}
\end{equation}

Thus, $\tilde{D}(\rho,\rho')$ approximates $\sum_l2\tilde{\varpi}_l$ and also gives us an approximate of the trace distance $D(\rho,\rho')$ between states $\rho$ and $\rho'$.


\section{Simulation and hardware results}\label{sec:simulation}

\begin{table*}[t]
\centering
\caption{Software simulation and IBM Brisbane results with different clock qubits of first QPE. PFD refers to the percentage fraction difference between simulation and hardware computations. }\label{table:hw results}

\begin{tabular}{*{10}{c}}
\hline \hline 
\multicolumn{1}{c}{Clock qubits}  &  \multicolumn{4}{c}{Simulation} & \multicolumn{4}{c}{Hardware} & \multicolumn{1}{c}{PFD} \\ 
\hline 
\multicolumn{1}{l}{} & 

\multicolumn{1}{c}{$\kappa_1$} & \multicolumn{1}{c}{$\kappa_2$} & \multicolumn{1}{c}{$\tilde{\ell}$} & \multicolumn{1}{c}{D} &

\multicolumn{1}{c}{$\kappa_1$} & \multicolumn{1}{c}{$\kappa_2$} & \multicolumn{1}{c}{$\tilde{\ell}$} & \multicolumn{1}{c}{D} &

\multicolumn{1}{c}{}
\\  
\multicolumn{1}{l}{$\tilde{D}(|0\rangle,|1\rangle)$} \\

2 & 0.078 & 0.031 & 0.000 & 1.000 & 0.078 & 0.033 & 0.048 & 1.013 & 1.30       
\\
3 & 0.078 & 0.031 & 0.000 & 1.000  & 0.079 & 0.034 & 0.041 & 1.002 & 0.20      
 \\ \hline

\\
\multicolumn{1}{l}{$\tilde{D}(|0\rangle,0.781|0\rangle+0.625|1\rangle)$}        & \multicolumn{1}{l}{}               & \multicolumn{1}{l}{}               & \multicolumn{1}{l}{} \\
2 & 0.058 & 0.035 & 0.960 & 0.897 & 0.057 & 0.044 & 0.983 & 1.055 & 17.61       \\
3 & 0.091 & 0.025 & 0.141 & 0.626 & 0.075 & 0.064 & 0.202 & 1.475 & 135.62       \\   
\hline

\\
\multicolumn{1}{l}{$\tilde{D}(|0\rangle,|0\rangle)$}          & \multicolumn{1}{l}{}               & \multicolumn{1}{l}{}               & \multicolumn{1}{l}{} \\
2 & 0.000 & 0.016 & 4.000 & 0.000 & 0.002 & 0.017 & 3.900 & 0.029 &  ---    \\
3 & 0.000 & 0.016 & 4.000 & 0.000 & 0.011 & 0.022 & 3.464 & 0.206 &  ---    \\ 
\hline \hline
\end{tabular}%

\end{table*}

In order to verify our algorithm, we simulate the first QPE in Fig.~\ref{fig:1}. For the sake of illustrations, we use Kitaev's QPE \cite{nielsen}. To verify our protocol, we utilize the exact matrix representation of the unitary $U_{AB}$ in the QPE, for example, we shall take $U_{AB} = e^{i (\rho-\rho')}$ instead of $e^{\rho}\cdot e^{-\rho'}$. Note that, as we are taking pure states in our numerical examples the unitary suffices to be $U_{AB} = e^{i (\rho-\rho')}$ instead of $U_{AB}= e^{i(\Omega/2)}$. We shall refer to the results obtained by simulating the first QPE on classical hardware with exact unitary as the simulation result, and it will include the theoretical error due to an insufficient number of clock qubits used in the QPE but will not include the Trotter error for the simulation of $U_{AB}$. 

\begin{figure}[!b]
    \centering
    \includegraphics[width=1\linewidth]{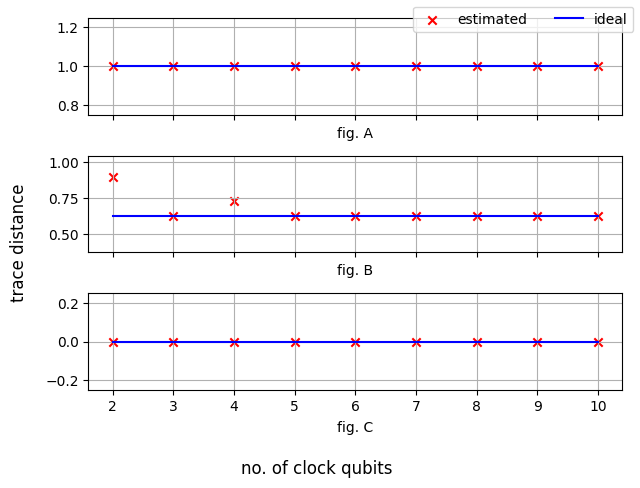}
    \caption{Plotting the software simulated trace distance value estimated by the algorithm containing only the precision error from the first QPE and comparing it with the ideal trace distance value that we aim to obtain.}
    \label{fig:sim-graph}
\end{figure}

We estimate the trace distances for three cases - $D(|0\rangle,|1\rangle)$, $D(|0\rangle,0.780|0\rangle + 0.625|1\rangle)$, and $D(|0\rangle,|0\rangle)$, and compare it with the ideal values of trace distance that can be obtained numerically. The results of this software simulation are visually represented by Fig.~\ref{fig:sim-graph}, and the exact values of the plotted points are in the Table \ref{table:software estimates}.

As is the characteristic of values obtained by phase estimation, these values also oscillate around the ideal value with an increasing number of clock qubits. Referring to Table~\ref{table:software estimates}, it is apparent that the difference between the estimated and ideal value can be either positive or negative. This results in the calculation of a trace distance that is slightly more than 1 in certain cases. However, it should be noted that, at a sufficiently large number of clock qubits the estimated trace distance value converges to the ideal value.

Another way to explain why the trace distance is estimated as greater than $1$ for insufficient clock qubits could be due to the way the formula for estimated trace distance (\ref{eq:kappa2-kappa1+l_estm}) is constructed. It consists of three variables - $\kappa_1$, $\kappa_2$ and $\tilde{\ell}$ which are individually estimated. In these individual circuits, overestimating $\kappa_2$ while underestimating $\kappa_1$ and $\tilde{\ell}$, due to precision error, could easily result in the overestimated values of trace distance.

We also simulate the first QPE with exact unitary on IBM's Brisbane processor, which in the context of this paper we shall call the hardware runs. The outcome of these hardware runs and their comparison with those simulated on a classical machine can be found in the Table~\ref{table:hw results}. 

Along with QPE precision error, the hardware also includes the errors inherent to current quantum hardware which could explain the percentage difference between the classical simulation and quantum hardware results.  

An important observation in the Table~\ref{table:hw results} is that we find the percentage difference for the case of $\tilde{D}(|0\rangle,|0\rangle)$ with $3$ clock qubits to be significantly higher than that of the other two cases of trace distance estimation with $3$ clock qubits. Rather than the greater depth of the circuit in this case, the percentage difference can be mostly attributed to the difference between the $\tilde{\ell}$ computed on classical hardware and quantum hardware. It appears that the formula is extremely sensitive to the estimated value of $\tilde{\ell}$ for $\tilde{D}(|0\rangle,|0\rangle)$.

Since the estimated $\tilde{D}(|0\rangle,|0\rangle)$ on hardware with $3$ clock qubits seems to be relatively computationally hard, we can precede the trace distance estimation algorithm with an algorithm that can tell us whether the two states are distinct. If they are distinct, we shall proceed to our trace distance estimation algorithm to quantify the distance between the two states, otherwise, halt if they are indistinct. One such state distinguishability method has been described in the discussion section \ref{sec:discussion}.


\begin{table}[h] 
\centering
\caption{Trace distance results upon running ideal simulations of the first QPE with 2 to 10 clock qubits on classical machine. }

\resizebox{\columnwidth}{!}{%
\begin{tabular}{cccc}
\hline\hline
\multicolumn{1}{l|}{~}  & ~ & software simulated estimates for  &  ~  \\ 
\multicolumn{1}{c|}{clock  qubits} & D = 1  & D = 0.625  & D = 0   \\ \hline
\multicolumn{1}{c|}{2}    & 1.000     & 0.897       & 0.000               \\
\multicolumn{1}{c|}{3}    & 1.000     & 0.626       & 0.000              \\
\multicolumn{1}{c|}{4}    & 1.000     & 0.733       & 0.000               \\ 
\multicolumn{1}{c|}{5}    & 1.000     & 0.625       & 0.000               \\ 
\multicolumn{1}{c|}{6}    & 1.000     & 0.625       & 0.000               \\
\multicolumn{1}{c|}{7}    & 1.000     & 0.625       & 0.000               \\ 
\multicolumn{1}{c|}{8}    & 1.000     & 0.625       & 0.000               \\
\multicolumn{1}{c|}{9}    & 1.000     & 0.625       & 0.000               \\
\multicolumn{1}{c|}{10}   & 1.000     & 0.625       & 0.000               \\ 
\hline\hline
\end{tabular}%
}
\label{table:software estimates}
\end{table}

\section{Complexity Analysis}\label{sec:complex}
\begin{itemize}
    \item \textbf{Complexity of simulating $U_{AB}$: }The complexity of simulating $U_{AB}$ from $U_A=e^{i\vartheta t}$ and $U_B=e^{-i\vartheta^\prime t}$ up to a small Trotter error $O(t^2)$ is $O(2(N+1)t^2/\varepsilon)$, where $t=O(1/\eta)$ owing to the subsequent improved quantum phase estimation. Here, $\varepsilon$ is the simulation error in trace distance of each of $U_A$ and $U_B$, and $\eta/2$ is the phase estimation precision error in trace distance. 
    \item \textbf{Complexity of simulating $Y$: }Similarly, the simulation of $Y=e^{i\beta t_1}$ has a complexity $O(M t_1^2/\delta)$, where $M$ is the number of qubits in the $|\tilde{\lambda}_j\rangle$ register, $\delta$ is the simulation error in trace distance, and $t_1=O(1/\zeta)$, such that $\zeta/2$ is the estimation precision error in trace distance.
    \item \textbf{Parallelization of the process: }Notice that the processes of obtaining the quantities $\kappa_1$ and $\kappa_2$ from $U_{AB}=e^{i (\Omega/2) t}$ and $U_{AB}=e^{i(\mathbb{I}+\Omega/2)t}$, respectively, can be carried out in parallel. 
    \item \textbf{Complexity of simulating $W$: }The simulation of $W=e^{i\gamma t_0}$ in each case has a complexity $O(t_0^2/\partial)$, where $\partial$ is the simulation error in trace distance, and $t_0=O(1/\upsilon)$, such that $\upsilon/2$ is the estimation precision error in trace distance. 
    \item \textbf{Overall complexity: }It is the complexity of these density matrix exponentiations that dominate in our algorithm. So, the overall complexity of our algorithm is $O(2(N+1)/(\varepsilon\eta^2)\times (M/(\delta\zeta^2)+1/(\partial\upsilon^2)))$. This is because the simulation of $U_{AB}$ (followed by phase estimation on it) needs to be repeated as many times as the number of copies of the state $\beta$ are required to simulate $Y$, and the simulation of $U_{AB}$ (followed by phase estimation on it, and controlled rotation of the ancilla qubit, conditioned on the phase estimates) needs to be repeated as many times as the number of copies $Q$ of the state $\gamma$ are required to simulate $W$. 
    \item \textbf{Simplifications: }Note that the quantities $\varepsilon$, $\eta$, $\delta$, $\zeta$, $\partial$ and $\upsilon$ determine the respective simulation or estimation errors in trace distance, and therefore, determine the maximum probabilities of errors of the corresponding simulations or estimations.
    Taking $\varepsilon=\eta=\delta=\zeta=\partial=\upsilon$ for simplicity, the overall complexity becomes $O(NM/\varepsilon^6)$, where $\varepsilon=O(1/{\rm poly}(N))$. Further taking $M=O(N)$ the overall complexity can be written as $O(N^2/\varepsilon^6)$. Ideally, we need $\varepsilon = O(1/2^N)$ to distinctly resolve all eigenvalues of $\Omega/2$, but if we relax the requirement of having to resolve all eigenvalues distinctly, we may have $\varepsilon = O(1/{\rm poly}(N))$, as long as the total cumulative error of the algorithm does not exceed 1/3. In that case, the overall complexity of our algorithm will be $O({\rm poly}(N))$, i.e. it will be efficient.
\end{itemize}
\section{Discussion}\label{sec:discussion}

The trace distance estimation algorithm described above can also be modified to distinguish between two arbitrary states $\rho$ and $\rho'$. For this, instead of the state in (\ref{eq:initial state}), create an equal superposition of states in the second register, to get the following initial state
\begin{equation}
    \frac{1}{\sqrt{2^N}}\sum_{j=0}^{2^N-1}{|\Psi_0\rangle|j\rangle}.
\end{equation}

Then we perform an IQPE with $U_{AB}=e^{i(\rho-\rho')t_1}$ to obtain the following state as the output:
\begin{equation}
    \frac{1}{\sqrt{2^N}}\sum_{j=0}^{2^N-1}{|\tilde{\varphi}\rangle|j\rangle}.
\end{equation}

Next, we trace out the second register and call the remaining state $\sigma$, given as,
\begin{equation}
    \sigma = \frac{1}{2^N}\sum_{j=0}^{2^N-1}{|\tilde{\varphi}\rangle\langle \tilde{\varphi}|}.
\end{equation}

If $\rho=\rho'$, then after the phase estimation, the eigenvalue for every state in the second register would be $1$ and the corresponding phase in the first register should be $0$. Additionally, it would also imply that after the trace operation for $\rho=\rho'$, $\sigma =|0\rangle\langle0|$.

Finally, the unitary $V=e^{i\sigma t_2}$ is created by employing density matrix exponentiation (DME), described in Ref.~\cite{LMR}. The unitary $V$ is used in another IQPE with input eigenstate $|0\rangle$ and time variable $t_2 = 1$. If the output phase estimate $\tilde{\phi}=1/(2\pi)$, then we can say $\rho=\rho'$ otherwise $\rho \neq \rho'$.

A popular method for determining whether two given pure states $|\psi\rangle$, $|\phi\rangle$ are equivalent is the SWAP-test~\cite{qfingerprinting}. It has a relatively simple circuit where the ancilla qubit on measurement outputs $1$ with the probability $(1-|\langle\psi|\phi\rangle|^2)/2$~\cite{dewolf2023quantum}. If the overlap is maximum, that is $|\langle\phi|\psi\rangle|=1$ , we get $1$ with probability $0$ but if the overlap $|\langle\phi|\psi\rangle|$ is very close to $0$, we obtain $1$ with a probability close to $1/2$.

Using SWAP-test, for the case where $|\psi\rangle \neq |\phi\rangle$ states, we can erroneously get all $0$s with the probability of atleast $(1/2)^{\text{shots}}$. This is contrasted by the state distinguishability method mentioned above, where a single shot of the algorithm has an error probability of at most $1/3$ for all cases of $|\psi\rangle$ and $|\phi\rangle$.

The SWAP-test can be extended to find the overlap of two mixed states~\cite{swaptestmixed}. However, among the two states being compared, if at most one is a mixed state we can estimate the value of state fidelity from the overlap values. In comparison, the above-proposed trace distance estimation algorithm works for any arbitrary quantum states.

Our algorithm faces errors arising from several distinct factors such as Trotterization inaccuracies occurring during the exponentiation of non-commuting density matrices within the LMR protocol, finite precision errors stemming from the limited number of clock qubits in the phase estimation stage, and general simulation errors alongside physical noise from the quantum hardware. The performance of our algorithm remains within the polynomial-time regime provided that the probability of error cumulated from all sources does not exceed 1/3. 

\section{Conclusion}\label{sec:con}
To summarize, we have designed an algorithm to estimate trace distance between two general quantum states applicable to both pure states as well as mixed states. By circumventing resource intensive procedure of quantum state tomography for estimating the quantum states, we have integrated LMR algorithm followed by IQPE to find out the sum of the absolute eigenvalues of the difference of the two states of interest. By avoiding full state construction, we can significantly reduce experimental overhead, providing a scalable framework. We have rigorously verified the efficacy of our algorithm by numerical simulation and hardware runs in an IBM processor, varying the number of clock qubits showing alignment with the ideal values. Our algorithm has an overall complexity of $O(N^2/\varepsilon^6)$ where N is the number of qubits of the states and $\varepsilon$ is the simulation or estimation precision error, offering a practical and robust solution for characterizing complex system and benchmarking quantum devices in NISQ era and future architecture.   
\section*{Acknowledgement}
We thank Akshaya Jayashankar and Srinivasa Prasannaa V.~for their valuable insights on this work. 
A.~B.~acknowledges the support received for this research from the research grant sanctioned by the National Board for Higher Mathematics (NBHM), Department of Atomic Energy (DAE), Government of India, 
with sanction letter no: 02011/32/2025/NBHM(R.P)/R\&D II/9677; the funding under the ARG program from Anusandhan National Research Foundation (ANRF) with 
file number: ANRF/ARG/2025/004617/PS; under the seed money scheme from Birla Institute of Technology Mesra with sanction letter no: DRIE/SMS/\\
DRIE-10917/2025-26/3857.

\bibliographystyle{ieeetr}
\bibliography{qsdbib}

\begin{thebibliography}{10}

\bibitem{nielsen}
M.~A. Nielsen and I.~L. Chuang, {\em Quantum computation and quantum information}.
\newblock Cambridge university press, 2010.

\bibitem{WZ}
W.~K. Wootters and W.~H. Zurek, ``A single quantum cannot be cloned,'' {\em Nature}, vol.~299, p.~802, October 1982.

\bibitem{DD}
D.~Dieks, ``Communication by {EPR} devices,'' {\em Physics Letters A}, vol.~92, p.~271, November 1982.

\bibitem{HPY}
H.~P. Yuen, ``Amplification of quantum states and noiseless photon amplifiers,'' {\em Physics Letters A}, vol.~113, p.~405, January 1986.

\bibitem{gil}
A.~Gilchrist, N.~K. Langford, and M.~A. Nielsen, ``Distance measures to compare real and ideal quantum processes,'' {\em Physical Review A}, vol.~71, p.~062310, June 2005.

\bibitem{BLP}
H.-P. Breuer, E.-M. Laine, and J.~Piilo, ``Measure for the degree of non-{M}arkovian behavior of quantum processes in open systems,'' {\em Physical Review Letters}, vol.~103, p.~210401, November 2009.

\bibitem{amato}
G.~Amato, H.-P. Breuer, and B.~Vacchini, ``Generalized trace distance approach to quantum non-{M}arkovianity and detection of initial correlations,'' {\em Physical Review A}, vol.~98, p.~012120, July 2018.

\bibitem{dajka}
J.~Dajka, J.~\L{}uczka, and P.~H\"anggi, ``Distance between quantum states in the presence of initial qubit-environment correlations: A comparative study,'' {\em Physical Review A}, vol.~84, p.~032120, September 2011.

\bibitem{wissman}
S.~Wi\ss{}mann, B.~Leggio, and H.-P. Breuer, ``Detecting initial system-environment correlations: Performance of various distance measures for quantum states,'' {\em Physical Review A}, vol.~88, p.~022108, August 2013.

\bibitem{qcorrelation}
A.~Smirne, D.~Brivio, S.~Cialdi, B.~Vacchini, and M.~G.~A. Paris, ``Experimental investigation of initial system-environment correlations via trace-distance evolution,'' {\em Physical Review A}, vol.~84, p.~032112, September 2011.

\bibitem{bell}
S.~G.~A. Brito, B.~Amaral, and R.~Chaves, ``Quantifying {B}ell nonlocality with the trace distance,'' {\em Physical Review A}, vol.~97, p.~022111, February 2018.

\bibitem{qcoherence}
Y.~Fan, X.~Guo, and X.~Yang, ``Quantifying coherence of quantum channels via trace distance,'' {\em Quantum Information Processing}, vol.~21, p.~339, September 2022.

\bibitem{Ai-Xi}
C.~Ai-Xi and L.~Jia-Hua, ``Effect of noise on trace distance of remote state preparation,'' {\em Chinese Physics}, vol.~14, p.~1507, august 2005.

\bibitem{Campbell}
J.~T.~M. Campbell, N.~Marchetti, J.~Dooley, and I.~Dey, ``{Testing Link Fidelity in a Quantum Network using Operational Form of Trace Distance with Error Bounds}.'' \arXiv{2404.10803}, April 2024.

\bibitem{network}
S.-H.~S. Pankaj~Kumar, Binayak~Kar, ``Trace-distance based end-to-end entanglement fidelity with information preservation in quantum networks,'' {\em Journal of Network and Computer Applications}, vol.~244, p.~104366, December 2025.

\bibitem{CWH1}
C.~W. Helstrom, ``Detection theory and quantum mechanics,'' {\em Information and Control}, vol.~10, p.~254, March 1967.

\bibitem{CWH2}
C.~W. Helstrom, ``Quantum detection and estimation theory,'' {\em Journal of Statistical Physics}, vol.~1, p.~231, June 1969.

\bibitem{AU}
A.~Uhlmann, ``The ``transition probability" in the state space of a $*$-algebra,'' {\em Reports on Mathematical Physics}, vol.~9, p.~273, April 1976.

\bibitem{LLSL}
S.-J. Li, J.-M. Liang, S.-Q. Shen, and M.~Li, ``Variational quantum algorithms for trace norms and their applications,'' {\em Communications in Theoretical Physics}, vol.~73, p.~105102, August 2021.

\bibitem{CSZW}
R.~Chen, Z.~Song, X.~Zhao, and X.~Wang, ``Variational quantum algorithms for trace distance and fidelity estimation,'' {\em Quantum Science and Technology}, vol.~7, p.~015019, December 2021.

\bibitem{WGLZY}
Q.~Wang, J.~Guan, J.~Liu, Z.~Zhang, and M.~Ying, ``New quantum algorithms for computing quantum entropies and distances,'' {\em IEEE Transactions on Information Theory}, vol.~70, pp.~5653--5680, August 2024.

\bibitem{WZ1}
Q.~Wang and Z.~Zhang, ``Fast quantum algorithms for trace distance estimation,'' {\em IEEE Transactions on Information Theory}, vol.~70, no.~4, pp.~2720--2733, 2023.

\bibitem{RASW}
S.~Rethinasamy, R.~Agarwal, K.~Sharma, and M.~M. Wilde, ``Estimating distinguishability measures on quantum computers,'' {\em Physical Review A}, vol.~108, p.~012409, July 2023.

\bibitem{qft}
J.~Zhang, P.~Ruggiero, and P.~Calabrese, ``Subsystem trace distance in quantum field theory,'' {\em Physical Review Letters}, vol.~122, p.~141602, April 2019.

\bibitem{spintomo}
M.~L. D\' Ariano G~M and P.~M, ``Spin tomography,'' {\em Journal of Optics B: Quantum and Semiclassical Optics}, vol.~5, p.~77–84, January 2003.

\bibitem{LMR}
S.~Lloyd, M.~Mohseni, and P.~Rebentrost, ``Quantum principal component analysis,'' {\em Nature Physics}, vol.~10, pp.~631--633, July 2014.

\bibitem{KSGS}
M.~Kjaergaard, M.~E. Schwartz, A.~Greene, G.~O. Samach, {\em et~al.}, ``Demonstration of density matrix exponentiation using a superconducting quantum processor,'' {\em Physical Review X}, vol.~12, p.~011005, January 2022.

\bibitem{BACS}
D.~W. Berry, G.~Ahokas, R.~Cleve, and B.~C. Sanders, ``Efficient quantum algorithms for simulating sparse {H}amiltonians,'' {\em Communications in Mathematical Physics}, vol.~270, pp.~359--371, March 2007.

\bibitem{HHL}
A.~W. Harrow, A.~Hassidim, and S.~Lloyd, ``Quantum algorithm for linear systems of equations,'' {\em Physical Review Letters}, vol.~103, p.~150502, October 2009.

\bibitem{qfingerprinting}
H.~Buhrman, R.~Cleve, J.~Watrous, and R.~De~Wolf, ``Quantum fingerprinting,'' {\em Physical Review Letters}, vol.~87, no.~16, p.~167902, 2001.

\bibitem{dewolf2023quantum}
R.~de~Wolf, ``Quantum computing: Lecture notes,'' 2023.

\bibitem{swaptestmixed}
H.~Kobayashi, K.~Matsumoto, and T.~Yamakami, ``Quantum {M}erlin-{A}rthur proof systems: Are multiple {M}erlins more helpful to {A}rthur?,'' in {\em Algorithms and Computation: 14th International Symposium, ISAAC 2003, Kyoto, Japan, December 15-17, 2003. Proceedings 14}, pp.~189--198, Springer, 2003.

\end{thebibliography}


\clearpage

\end{document}